# A new meshless approach for three dimensional fluid flow and related heat transfer problems

Cheng-An Wang, Hamou Sadat and Christian Prax

Institut PPRIME - CNRS - Université de Poitiers – ENSMA, Département Fluides, Thermique, Combustion - Campus Sud 40, avenue du recteur Pineau - 86022 POITIERS Cedex - France

**Abstract:**

The mathematical formulation, basic concept and numerical implementation of a new meshless method for solving three dimensional fluid flow and related heat transfer problems are presented in this paper. Moving least squares approximation is used for the spatial discretization together with an implicit scheme for time marching. The vorticity and vector potential formulation of Navier-Stokes equations is employed to avoid the difficulties associated with pressure-velocity coupling. Two three–dimensional problem of natural convection in a differentially heated cubic cavity and in the annular space between a sphere and a cube are considered. Results show the flexibility of the method and demonstrate its accuracy.

## 1 Introduction

Among the numerical methods available for solving partial differential equations, meshless methods are gaining growing interest. Many different approaches (collocation, Galerkin, Petrov-Galerkin...) have been proposed so far and a number of textbooks have already been published [1]-[4]. These methods have been applied by many researchers for solving fluid flow and heat transfer problems [5]-[13]. In this work, we will focus on the diffuse approximation based meshless method which has been used for the first time in [14] for solving an annular convection flow. This method was further extended to deal with natural convection in porous media [15] and with the flow in a two dimensional channel[16]. A comparison with the control volume based finite element method has been presented in [17] and has shown the good accuracy of the method. These first two-dimensional works were done by employing the streamfunction-vorticity formulation of the Navier Stokes equations. The method has been extended to the primitive variables formulation in [18] by using a projection algorithm to deal with the pressure velocity coupling. Results obtained therein in channel and cavity flows were very encouraging. A more complex geometry has been considered in [19] where natural convection in an annular space between an inner hot cylinder and an external cold square cavity has been studied. The ability of the method to tackle unsteady problems has been examined in [20] where the wake flow over a circular and semi-circular cylinder has been numerically calculated. The Von Karmann street was put in evidence and the Strouhal numbers calculated. It is worth noting here that the method has been used for solving an international benchmark problem of convection dominating phase change problem. The method was one of the two more accurate [21]. The unsteady pseudo periodic natural convection flow in a two dimensional differentially heated enclosure has been examined and reported in [22] where the efficiency of the method was demonstrated. The extension to three dimensional natural convection was finally done in [23] by considering the differentially heated cavity up to a Rayleigh number of $Ra=10^6$. Although the results were of very good accuracy, we have observed that the pressure equation obtained when using the projection algorithm was very badly conditioned in some situations leading to the increase of the CPU time. The main goal of this work is therefore to present a vorticity-vector potential formulation based meshless method to solve natural convection problem in a three dimensional complex geometry. In the following sections, the meshless approximation is rapidly presented. The mathematical formulation of the problem and the discretization schemes are then given. Finally two cases are solved to assess the correctness of the present numerical approach.

## 2   The diffuse approximation method

The approximation used herein is the Diffuse Approximation Method (DAM) presented by Nayroles et al. [24][25]. This approach is very closely related to the moving least squares approximation developed earlier by Lancaster et al. [26] for surface generation. Considering a scalar three dimensional field $\varphi$, at each nodal point $X$ of the discretization, the Taylor expansion of $\varphi$ is estimated by a weighted least squares method which uses only the values of $\varphi$ at some points $X_i$ situated in the vicinity of $X$:

$$\varphi_i^{estimated} = \langle P(X_i - X) \rangle \langle \alpha(X) \rangle^T \qquad (1)$$

where $\langle P(X_i - X) \rangle$ is a vector of polynomial basis functions and $\langle \alpha(X) \rangle^T$ a vector of coefficients which are determined by minimizing the functional:

$$I(\alpha) = \sum_{i=1}^{N} \omega(X, X_i - X) \left[ \varphi_i - \langle P(X_i - X) \rangle \langle \alpha(X) \rangle^T \right]^2 \qquad (2)$$

in which $\omega(X, X_i - X)$ is a weight-function of compact support, equal to unity at this point and decreasing when the distance to the node increases. Its value is set to zero outside a given domain of influence (a more precise description of $\omega(X, X_i - X)$ will be done next). Minimization of (2) then gives:

$$A(X)\alpha(X) = B(X) \qquad (3)$$

where:

$$A(X) = \sum_{i=1}^{N} \omega(X, X_i - X) P(X_i - X) P^T(X_i - X) \qquad (4)$$

$$B(X) = \sum_{i=1}^{N} \omega(X, X_i - X) P(X_i - X) \varphi_i \qquad (5)$$

The Taylor expansion can be truncated at any order $k$ giving rise to different order approximations [27]. In this work, the Taylor expansion is truncated at order 2 and the polynomial vector is:

$$\langle p(X_i, X) \rangle = \langle 1, (x_i - x), (y_i - y), (x_i - x)^2, (x_i - x)(y_i - y), (y_i - y)^2,$$
$$(z_i - z), (x_i - x)(z_i - z), (y_i - y)(z_i - z), (z_i - z)^2 \rangle \qquad (6)$$

The components of generalized variables vector $\alpha(X)$ are therefore given therefore by:

$$\alpha_1 = \varphi(x,y,z)^* \quad \alpha_2 = \left(\frac{\partial \varphi}{\partial x}\right)^* \quad \alpha_3 = \left(\frac{\partial \varphi}{\partial y}\right)^* \quad \alpha_4 = \frac{1}{2!}\left(\frac{\partial^2 \varphi}{\partial x^2}\right)^* \quad \alpha_5 = \left(\frac{\partial^2 \varphi}{\partial x \partial y}\right)^*$$
$$\alpha_6 = \frac{1}{2!}\left(\frac{\partial^2 \varphi}{\partial y^2}\right)^* \quad \alpha_7 = \left(\frac{\partial \varphi}{\partial z}\right)^* \quad \alpha_8 = \left(\frac{\partial^2 \varphi}{\partial x \partial z}\right)^* \quad \alpha_9 = \left(\frac{\partial^2 \varphi}{\partial y \partial z}\right)^* \quad \alpha_{10} = \frac{1}{2!}\left(\frac{\partial^2 \varphi}{\partial z^2}\right)^* \qquad (7)$$

By inverting system (3), one obtains the components of $\alpha$ which are the estimated derivatives of $\varphi$ at $X$ in terms of the neighboring nodal values $\varphi_i$:

$$\left[ \varphi \quad \frac{\partial \varphi}{\partial x} \quad \frac{\partial \varphi}{\partial y} \quad \frac{\partial^2 \varphi}{2! \partial x^2} \quad \frac{\partial^2 \varphi}{\partial x \partial y} \quad \frac{\partial^2 \varphi}{2! \partial y^2} \quad \frac{\partial \varphi}{\partial z} \quad \frac{\partial^2 \varphi}{\partial x \partial z} \quad \frac{\partial^2 \varphi}{\partial z \partial y} \quad \frac{\partial^2 \varphi}{2! \partial z^2} \right]^{*T}$$
$$= [A(X)]^{-1} \cdot \left\{ \sum_{i=1}^{n'(X)} \omega(X, X_i - X) \langle P(X_i - X) \rangle^T \cdot \varphi_i \right\} \qquad (8)$$

Note that the square matrix $A(X)$ is not singular as long as the number $n(X)$ of the connected nodes at a given point is at least equal to the size of $\langle P(X_i - X) \rangle$ and are not all situated in the same plane. Finally, the following Gaussian window:

$$\omega(M_j, M) = \exp\left[-3\ln(10) \cdot \left(\frac{r}{\sigma}\right)^2\right] \quad (9)$$

$$\omega(M_j, M) = 0, \text{ if } r > \sigma \quad (10)$$

has been used in the calculations, the distance of influence $\sigma$ being updated at each calculation point.

The previous approximation is now simply used in a point collocation method to solve partial derivatives equations. At each point of the nodes set, the derivatives appearing in the equation to be solved are replaced by their diffuse approximation thus leading to an algebraic system that is solved after the introduction of the boundary conditions. Dirichlet type boundary conditions are introduced in the same way as in the finite element method. Neumann boundary conditions on the other hand are replaced by their diffuse approximation and then introduced in the algebraic system.

## 3 Vorticity and Vector Potential Formulation of Navier Stokes equations

We give in this section the governing adimensional equations of the 3D natural convection problem within the framework of the vorticity and vector potential formulation. The Boussinesq approximation which states that density depends weakly on temperature has been used. We define $\Delta T = T_h - T_c$ and $T_{ref} = (T_h + T_c)/2$ where $T_h$ and $T_c$ are the high and low temperatures and we introduce the following dimensionless variables:

$$\vec{V} = \frac{\vec{V}^+}{V_0}, \quad X = \frac{X^+}{L}, \quad \vec{\Omega} = \frac{\vec{\Omega}^+ L}{V_0}, \quad \tau = \frac{vt}{L^2}, \quad T = \frac{T^+ - T_{ref}}{\Delta T}, \quad V_0 = \frac{\alpha}{L}$$

where $\vec{V}$ and $\vec{\Omega}$ are the velocity and vorticity vectors, $\alpha$ and $L$ are the thermal diffusivity and a reference dimension and $v$ and $t$ are the kinematic viscosity and time.

From the equation of continuity of Navier Stokes equations ($\vec{\nabla} \cdot \vec{V} = 0$) it follows that the velocity is a solenoidal field. Thus the velocity, can be determined by a vector function $\vec{A}$ which is defined as: $\vec{V} = \vec{\nabla} \wedge \vec{A}$. Let us suppose further that the vector function satisfies the additional condition $\vec{\nabla} \cdot \vec{A} = 0$. Introducing the vorticity as $\vec{\Omega} = \vec{\nabla} \wedge \vec{V}$, one gets the following relation between the vector potential and the vorticity vector:

$$\vec{\Omega} = -\nabla^2 \vec{A} \quad (11)$$

Hence the vector potential can be obtained from the vorticity by solving the three Poisson equations:

$$\nabla^2 A_x = -\xi \quad (12)$$

$$\nabla^2 A_y = -\eta \quad (13)$$

$$\nabla^2 A_z = -\zeta \quad (14)$$

If we now apply the operator ($\vec{\nabla} \wedge$) to the momentum equations, we obtain the following governing equations for the vorticity components:

$$\frac{\partial \xi}{\partial t} + u\frac{\partial \xi}{\partial x} + v\frac{\partial \xi}{\partial y} + w\frac{\partial \xi}{\partial z} = \xi\frac{\partial u}{\partial x} + \eta\frac{\partial u}{\partial y} + \zeta\frac{\partial u}{\partial z} + Pr\nabla^2\xi + RaPr\left(\frac{\partial T}{\partial y}\right) \quad (15)$$

$$\frac{\partial \eta}{\partial t} + u\frac{\partial \eta}{\partial x} + v\frac{\partial \eta}{\partial y} + w\frac{\partial \eta}{\partial z} = \xi\frac{\partial v}{\partial x} + \eta\frac{\partial v}{\partial y} + \zeta\frac{\partial v}{\partial z} + Pr\nabla^2\eta - RaPr\frac{\partial T}{\partial x} \quad (16)$$

$$\frac{\partial \zeta}{\partial t} + u\frac{\partial \zeta}{\partial x} + v\frac{\partial \zeta}{\partial y} + w\frac{\partial \zeta}{\partial z} = \xi\frac{\partial w}{\partial x} + \eta\frac{\partial w}{\partial y} + \zeta\frac{\partial w}{\partial z} + Pr\nabla^2\zeta \quad (17)$$

Where $Pr = \nu/\alpha$ and $Ra = \dfrac{g\beta\Delta T L^3}{\nu\alpha}$ are the Prandtl and Rayleigh numbers respectively. This form of the momentum equation with the velocity vector contains no pressure term. The energy equation is the last equation closing the whole formulation:

$$\frac{\partial T}{\partial t} + u\frac{\partial T}{\partial x} + v\frac{\partial T}{\partial y} + w\frac{\partial T}{\partial z} = \nabla^2 T \quad (18)$$

We now have a system of seven equations with seven unknowns which is solved iteratively after the introduction of the appropriate boundary conditions.

### 3.1 Boundary conditions

At all walls velocity is set to zero (no slip condition) and vorticity is calculated by using the relation $\vec{\Omega} = \vec{\nabla} \wedge \vec{V}$. The boundary conditions for the vector potential can be used in the form: $\dfrac{\partial \psi_n}{\partial n} = \psi_t = \psi_q = 0$ where $n$, $t$ and $q$ denote the normal and the two tangential components.

### 3.2 Simulation algorithm

The iterative algorithm is as follows:

1. An irregular set of $N$ nodes is first positioned in the calculation domain.

2. Fields are initialized : $\xi^0, \eta^0, \zeta^0, u^0, v^0, w^0, T^0$

3. Solve the equations for the new components of the vector potential:

$$\nabla^2 \psi_x^{t+1} = -\xi^t \tag{19}$$

$$\nabla^2 \psi_y^{t+1} = -\eta^t \tag{20}$$

$$\nabla^2 \psi_z^{t+1} = -\zeta^t \tag{21}$$

4. Update velocity components:

$$u^{t+1} = \frac{\partial \psi_z^{t+1}}{\partial y} - \frac{\partial \psi_y^{t+1}}{\partial z} \tag{22}$$

$$v^{t+1} = \frac{\partial \psi_x^{t+1}}{\partial z} - \frac{\partial \psi_z^{t+1}}{\partial x} \tag{23}$$

$$w^{t+1} = \frac{\partial \psi_y^{t+1}}{\partial x} - \frac{\partial \psi_x^{t+1}}{\partial y} \tag{24}$$

5. Solve vorticity transport equations:

$$\frac{\partial \xi^{t+1}}{\partial t} + u^t \frac{\partial \xi^{t+1}}{\partial x} + v^t \frac{\partial \xi^{t+1}}{\partial y} + w^t \frac{\partial \xi^{t+1}}{\partial z} = \xi^t \frac{\partial u^t}{\partial x} + \eta^t \frac{\partial u^t}{\partial y} + \zeta^t \frac{\partial u^t}{\partial z} \\ + Pr\nabla^2 \xi^t + RaPr\left(\frac{\partial T^t}{\partial y}\right) \tag{25}$$

$$\frac{\partial \eta^{t+1}}{\partial t} + u^t \frac{\partial \eta^{t+1}}{\partial x} + v^t \frac{\partial \eta^{t+1}}{\partial y} + w^t \frac{\partial \eta^{t+1}}{\partial z} = \xi^t \frac{\partial v^t}{\partial x} + \eta^t \frac{\partial v^t}{\partial y} + \zeta^t \frac{\partial v^t}{\partial z} \\ + Pr\nabla^2 \eta^t - RaPr\frac{\partial T^t}{\partial x} \tag{26}$$

$$\frac{\partial \zeta^{t+1}}{\partial t} + u^t \frac{\partial \zeta^{t+1}}{\partial x} + v^t \frac{\partial \zeta^{t+1}}{\partial y} + w^t \frac{\partial \zeta^{t+1}}{\partial z} = \xi^t \frac{\partial w^t}{\partial x} + \eta^t \frac{\partial w^t}{\partial y} + \zeta^t \frac{\partial w^t}{\partial z} \\ + Pr\nabla^2 \zeta^t \tag{27}$$

6. Solve energy equation :

$$\frac{\partial T^{t+1}}{\partial t} + u^t \frac{\partial T^{t+1}}{\partial x} + v^t \frac{\partial T^{t+1}}{\partial y} + w^t \frac{\partial T^{t+1}}{\partial z} = \nabla^2 T^t \tag{28}$$

7. Stop the iteration process if the convergence criterion is satisfied. Otherwise, go back to step 3.

The convergence criterion used in the numerical calculation for all the variables is the following:

$$\left|\frac{\varphi_{new} - \varphi_{old}}{\varphi_{new}}\right|_{max} \leq 10^{-5}$$

# 4 Discretization schemes

The seven algebraic systems can all be put in the general form:

$$[MAT]_Z \times [Z] = [S]_Z \qquad (29)$$

Where $MAT$ is the $N \times N$ matrix and $S$ the $N$ second member vector. $Z$ stands for the unknown and represents respectively one of the following variables: $[A_x, A_y, A_z, \xi, \eta, \zeta, T]$.

## 4.1 Vector potential equations : $Z \in [A_x, A_y, A_z]$.

By writing $\langle p(X_j, X) \rangle = \langle p_j \rangle$ and if $\langle a_i \rangle$ is the $i^{th}$ line of inverse matrix $[A^M]^{-1}$, we have the following expression for the $i^{th}$ line of matrix $[MAT]_Z$:

$$MAT_Z(i,j) = \omega(X_j, X) \cdot [2!\langle a_4 \rangle + 2!\langle a_6 \rangle + 2!\langle a_{10} \rangle] \cdot \langle p_j \rangle^T \qquad (30)$$

The second members write respectively:

$$S_{A_x}(i) = -\xi^t(i) \qquad (31)$$
$$S_{A_y}(i) = -\eta^t(i) \qquad (32)$$
$$S_{A_z}(i) = -\zeta^t(i) \qquad (33)$$

## 4.2 Velocity Components $u, v, w$ :

Once the vectoral potential calculated, the velocity is updated as follows:

$$u_i^{t+1} = \sum_{M_j \in \upsilon^M} \omega(M_j, M) \cdot (\langle a_3 \rangle) \cdot \langle P_j \rangle^T \cdot \psi_z^{t+1}(j) - \sum_{M_j \in \upsilon^M} \omega(M_j, M) \cdot (\langle a_7 \rangle) \cdot \langle P_j \rangle^T \cdot \psi_y^{t+1}(j) \qquad (34)$$

$$v_i^{t+1} = \sum_{M_j \in \upsilon^M} \omega(M_j, M) \cdot (\langle a_7 \rangle) \cdot \langle P_j \rangle^T \cdot \psi_x^{t+1}(j) - \sum_{M_j \in \upsilon^M} \omega(M_j, M) \cdot (\langle a_2 \rangle) \cdot \langle P_j \rangle^T \cdot \psi_z^{t+1}(j) \qquad (35)$$

$$w_i^{t+1} = \sum_{M_j \in \upsilon^M} \omega(M_j, M) \cdot (\langle a_2 \rangle) \cdot \langle P_j \rangle^T \cdot \psi_y^{t+1}(j) - \sum_{M_j \in \upsilon^M} \omega(M_j, M) \cdot (\langle a_3 \rangle) \cdot \langle P_j \rangle^T \cdot \psi_x^{t+1}(j) \qquad (36)$$

## 4.3 Vorticity and temperature equations : $Z \in [\xi, \eta, \zeta, T]$.

The the $i^{th}$ line of matrix $[MAT]_Z$ is now:

$$MAT_Z(i,j) = \omega(M_j, M) \cdot$$
$$[\langle a_1 \rangle / \Delta t + \langle a_2 \rangle \cdot u^t(j) + \langle a_3 \rangle \cdot v^t(j) + \langle a_7 \rangle \cdot w^t(j) - \Pr(2!\langle a_4 \rangle + 2!\langle a_6 \rangle + 2!\langle a_{10} \rangle)] \qquad (37)$$
$$\cdot \langle p_j \rangle^T$$

The second members write respectively :

$$S_\xi(i) = \sum_{M_j \in \upsilon^M} \omega(M_j, M) \cdot \left(\xi^t(i) \cdot \langle a_2 \rangle + \eta^t(i) \cdot \langle a_3 \rangle + \zeta^t(i) \cdot \langle a_7 \rangle\right) \cdot \langle P_j \rangle^T \cdot u^t(j)$$
$$+ Ra \Pr \sum_{M_j \in \upsilon^M} \omega(M_j, M) \cdot (\langle a_3 \rangle) \cdot \langle P_j \rangle^T \cdot T^t(j) + \xi^t(i)/\Delta t \tag{38}$$

$$S_\eta(i) = \sum_{M_j \in \upsilon^M} \omega(M_j, M) \cdot \left(\xi^t(i) \cdot \langle a_2 \rangle + \eta^t(i) \cdot \langle a_3 \rangle + \zeta^t(i) \cdot \langle a_7 \rangle\right) \cdot \langle P_j \rangle^T \cdot v^t(j)$$
$$+ Ra \Pr \sum_{M_j \in \upsilon^M} \omega(M_j, M) \cdot (-\langle a_2 \rangle) \cdot \langle P_j \rangle^T \cdot T^t(j) + \eta^t(i)/\Delta t \tag{39}$$

$$S_\zeta(i) = \sum_{M_j \in \upsilon^M} \omega(M_j, M) \cdot \left(\xi^t(i) \cdot \langle a_2 \rangle + \eta^t(i) \cdot \langle a_3 \rangle + \zeta^t(i) \cdot \langle a_7 \rangle\right) \cdot \langle P_j \rangle^T \cdot w^t(j)$$
$$+ \zeta^t(i)/\Delta t \tag{40}$$

$$S_T(i) = T^t(i)/\Delta t \tag{41}$$

### 4.4 Boundary conditions

Dirichlet Boundary conditions can be introduced as in the finite element method by using the unit term on the diagonal for example. They can also be introduced by using the first line of system (8). Appropriate lines of system (8) are used for the introduction of Neumann boundary as well. As a matter of example, the Dirichlet boundary condition for temperature can be introduced by writing:

$$MAT_T(i,j) = \omega(M_j, M) \cdot [\langle a_1 \rangle] \cdot \langle p_j \rangle^T \tag{42}$$
$$S_T(i) = T_w(i) \tag{43}$$

For the homogeneous Neumann type boundary condition we write:

$$MAT_T(i,j) = \omega(M_j, M) \cdot [n_x \langle a_2 \rangle + n_y \langle a_3 \rangle + n_z \langle a_7 \rangle] \cdot \langle p_j \rangle^T \tag{44}$$
$$S_T(i) = 0 \tag{45}$$

## 5 Results and discussion

In this section we present the results obtained by the present method when applied to two natural convection problems. The first one is the differentially heated cubic cavity and the second one is related to convection in the annular space between a sphere and a cavity. In each case, a non uniform cloud of nodes is generated in the cavity with a grid refinement near the solid boundaries. A grid sensivity of the solution has been carried out in both cases.

## 5.1 Differentially heated cubic cavity

We consider here the differentially heated unit cubic cavity shown on Figure 1. Surfaces at $x=0$ and $x=1$ are maintained at temperatures $T_h$ and $T_c$ respectively while the remaining surfaces are considered adiabatic. Prandtl number was set to $Pr=0.71$ and Rayleigh numbers ranging from $10^3$ to $10^6$ have been considered. We give on Figure 2 the isotherms obtained for different Rayleigh numbers and on Figure 3 the isovorticities in the plane $y=0.5$. The results are in good agreement with the available literature results [28]. On Figure 4, are gathered the mean Nusselt numbers along the $y$ direction defined by:

$$Nu_{mean} = \int_0^1 -\frac{\partial T}{\partial x} dz$$

As can be seen, we have obtained the same results as those given by [28]. Finally, The Average Nusselt number $\overline{Nu} = \iint -\frac{\partial T}{\partial x} dydz$ at the surface $x=0$ is given on table 1 where the results of [29] and [30] are also presented. The calculated Nusselt numbers are very close to the reference values with a maximum relative error lower than 1.5%.

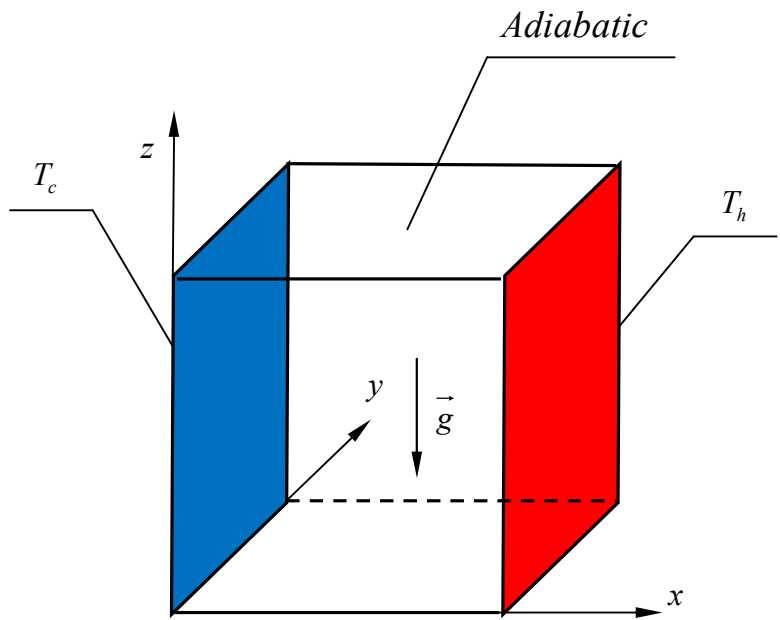

Figure 1 : Scheme of the cubic enclosure

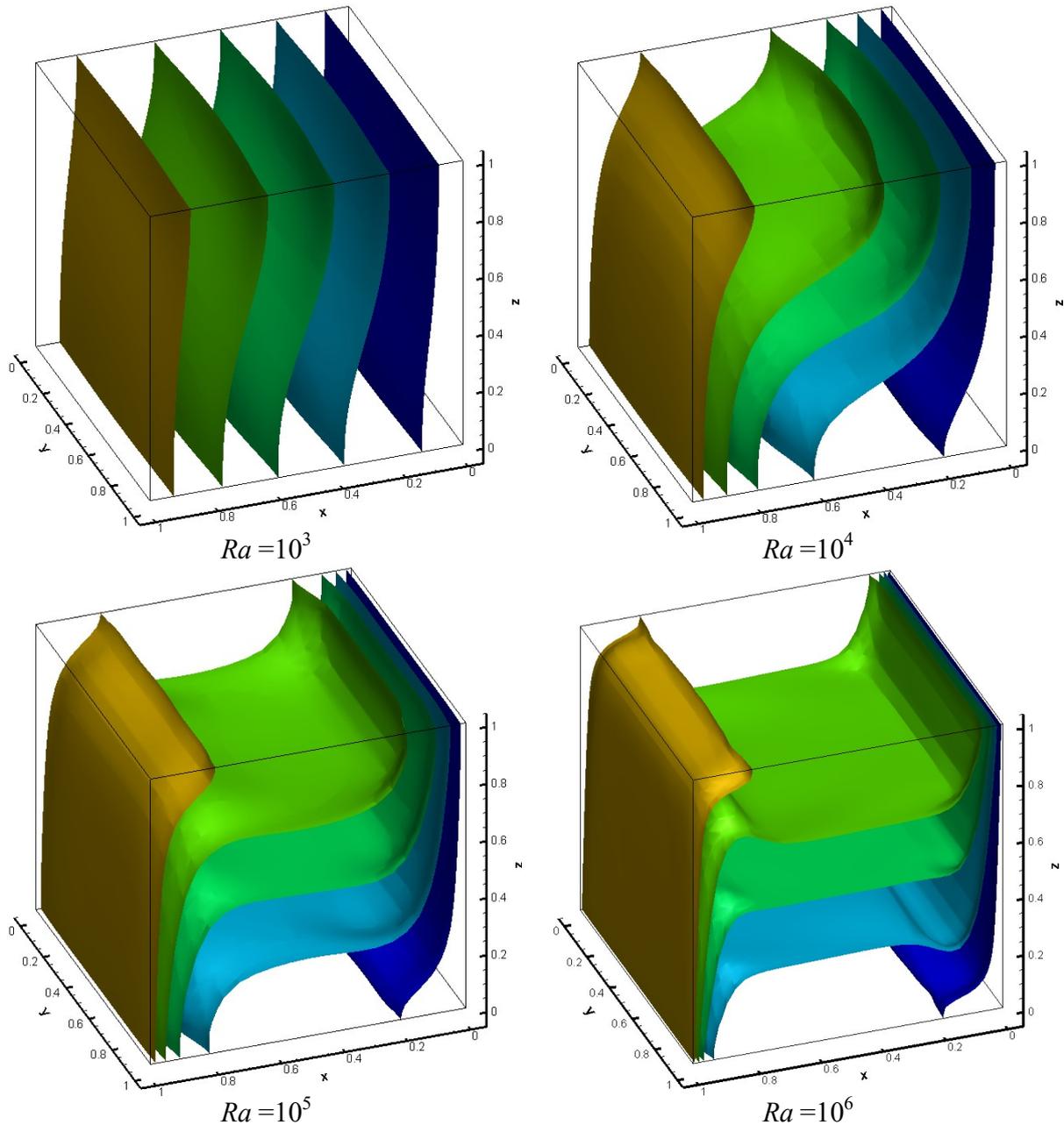

Figure 2 : Isotherms for different Rayleigh numbers

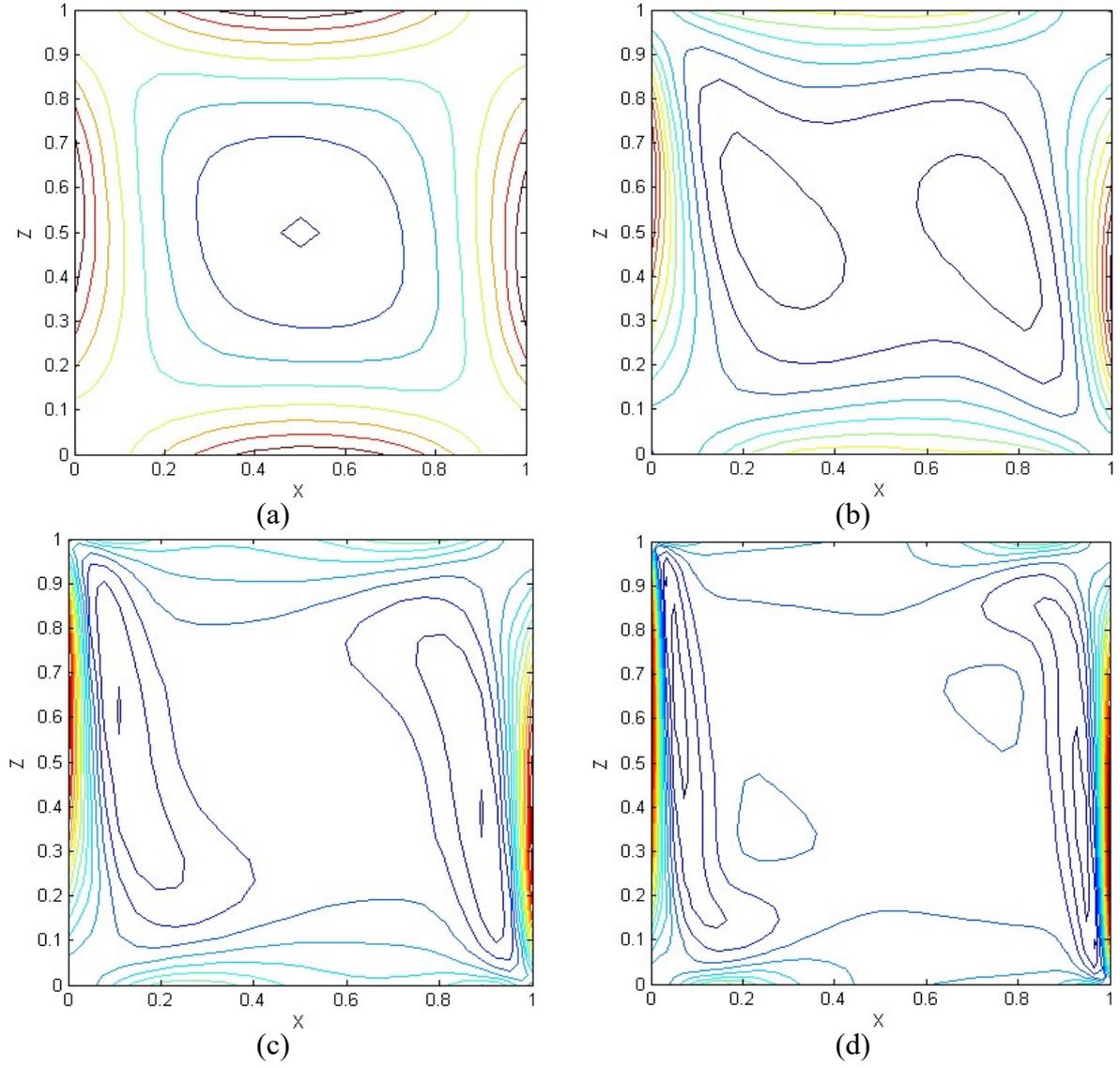

Figure 3 : Vorticity for $Ra = 10^3$ (a), $10^4$ (b), $Ra = 10^5$ (c) and $10^6$ (d)

|  | Grid | $\overline{Nu}$ | PSC [29] | FDM [30] |
|---|---|---|---|---|
| $Ra = 10^3$ | 25×25×25 | 1.06966 | 1.0700 | 1.085 |
| $Ra = 10^4$ | 25×25×25 | 2.07029 | 2.0542 | 2.100 |
|  | 31×31×31 | 2.06385 |  |  |
| $Ra = 10^5$ | 25×25×25 | 4.42922 | 4.3371 | 4.361 |
|  | 31×31×31 | 4.39224 |  |  |
|  | 35×35×35 | 4.38253 |  |  |
| $Ra = 10^6$ | 31×31×31 | 9.00744 | 8.6407 | 8.770 |
|  | 35×35×35 | 8.93559 |  |  |
|  | 41×41×41 | 8.82119 |  |  |
|  | 51×51×51 | 8.77027 |  |  |

Table 1: Average Nusselt number

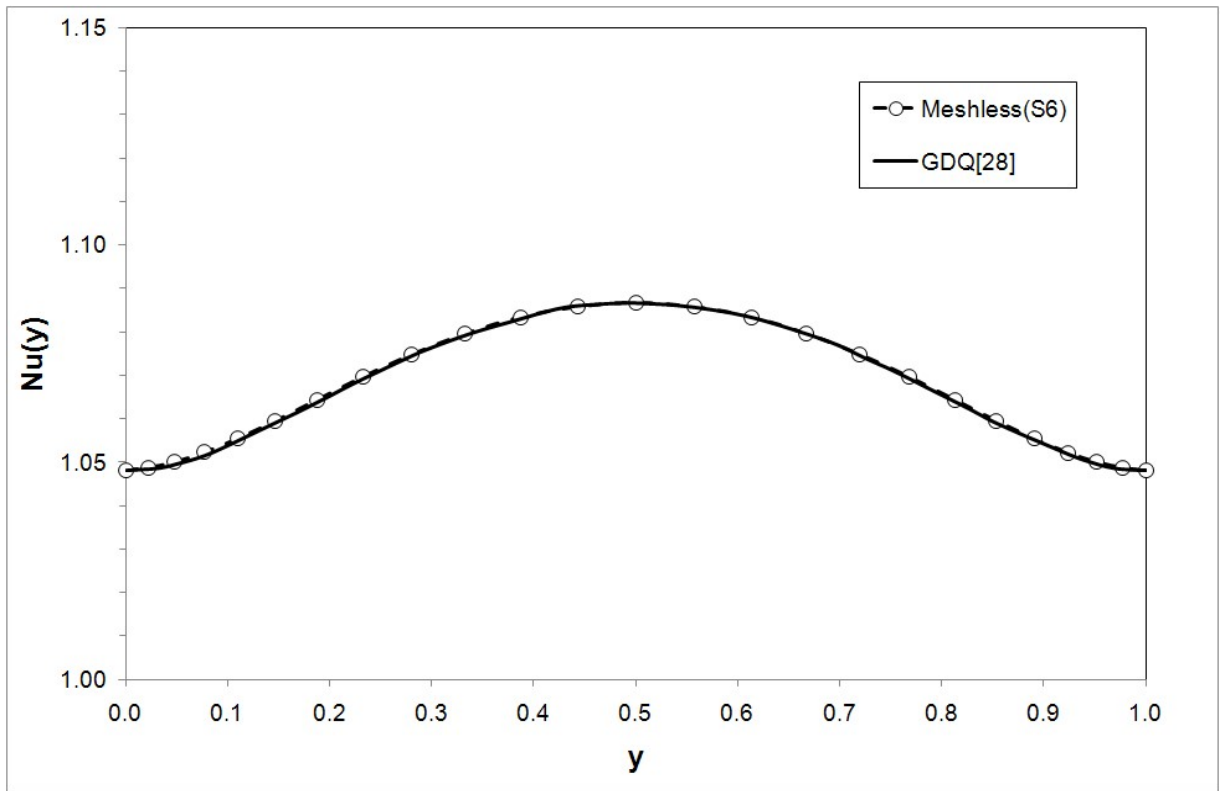

(a)

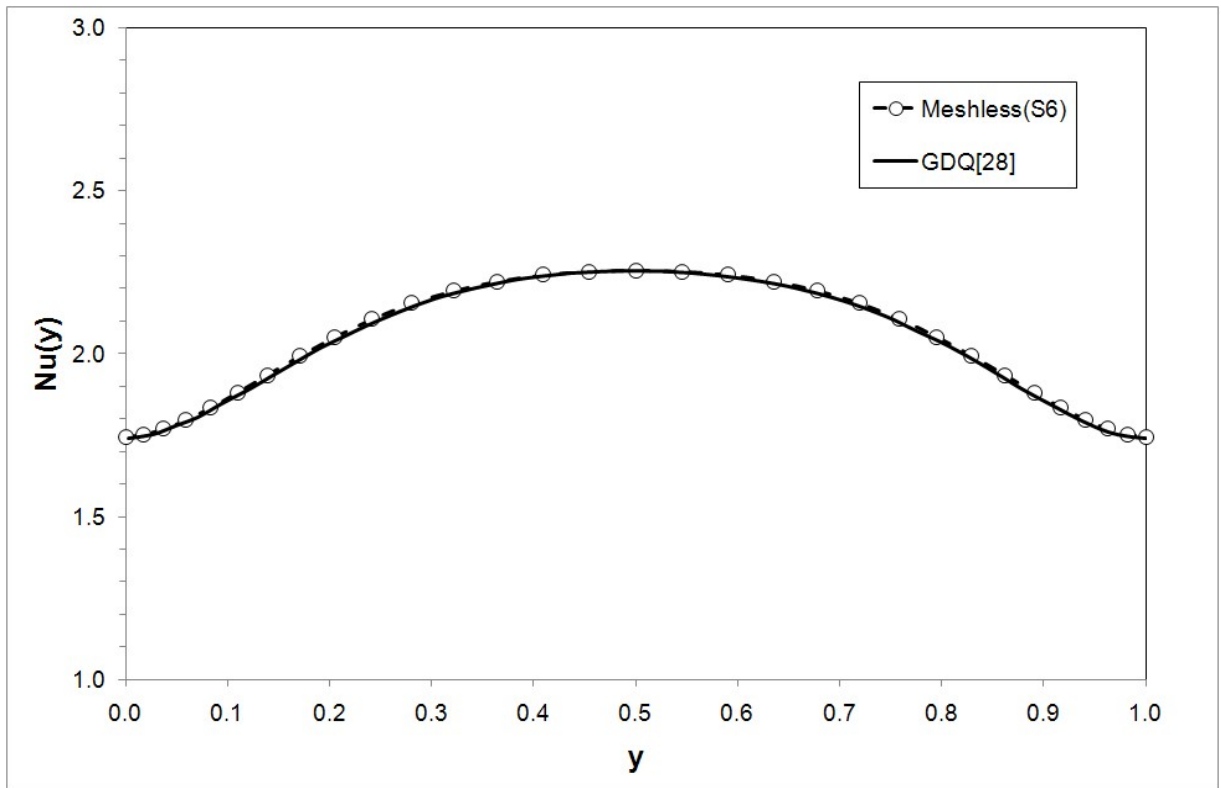

(b)

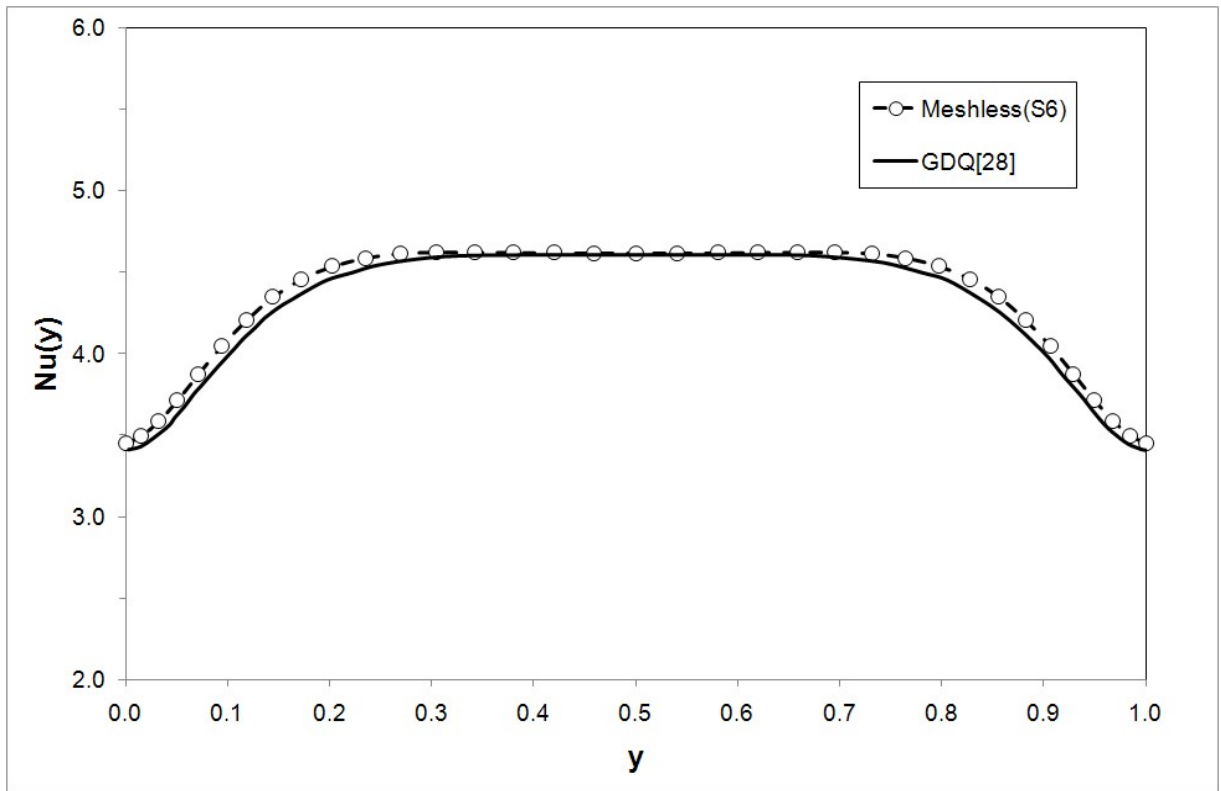

(c)

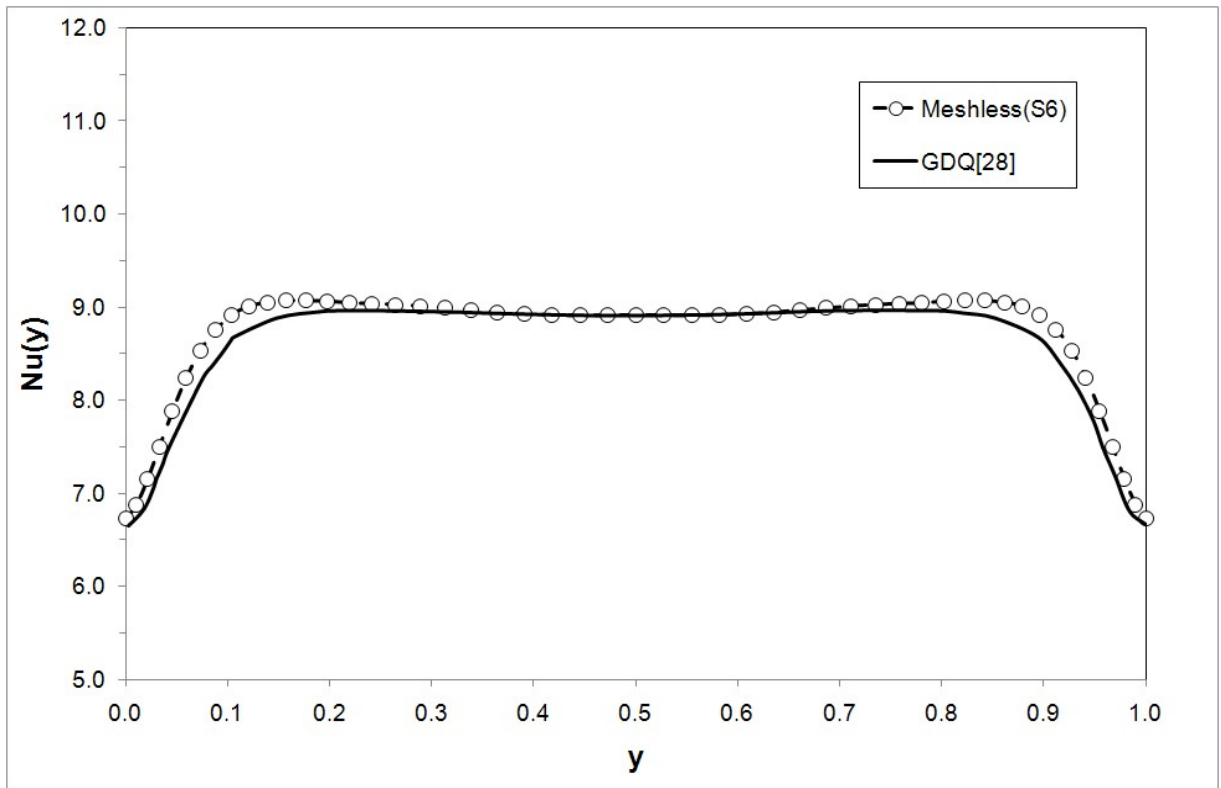

(d)

Figure 4 : Mean Nusselt number for $Ra = 10^3$ (a), $10^4$ (b), $10^5$ (c), $10^6$ (d)

## 5.2 Annular space between a sphere and a cubic cavity

As a second illustrative example, we consider the case of a hot sphere at temperature $T_h$ placed in a cold cubic cavity of length $L$ maintained at temperature $T_c$ (Figure 5). The tolerance error has been set to $10^{-5}$ in the calculations. This problem can be viewed as the three dimensional counterpart of the two dimensional case of a hot cylinder placed in an cold enclosure already treated in [19] by a meshless method coupled to a projection algorithm
The problem has been solved by Yoon et al. [31] who used an immersed boundary finite volume method. We therefore will present as in [31], the results in the two planes evidenced in Figure 5 by doted lines and corresponding to $\theta=0$ and $\theta=45°$.
The present results have been obtained with a set of 67934 nodes (41×41 points in the square boundary of the cavity) for $Ra =10^3, 10^4$ and $10^5$. For $Ra =10^6$ the number of nodes was increased to a total of 129636 nodes (51×51 in the square surface of the cavity). Typical grid is shown on Figure 6.

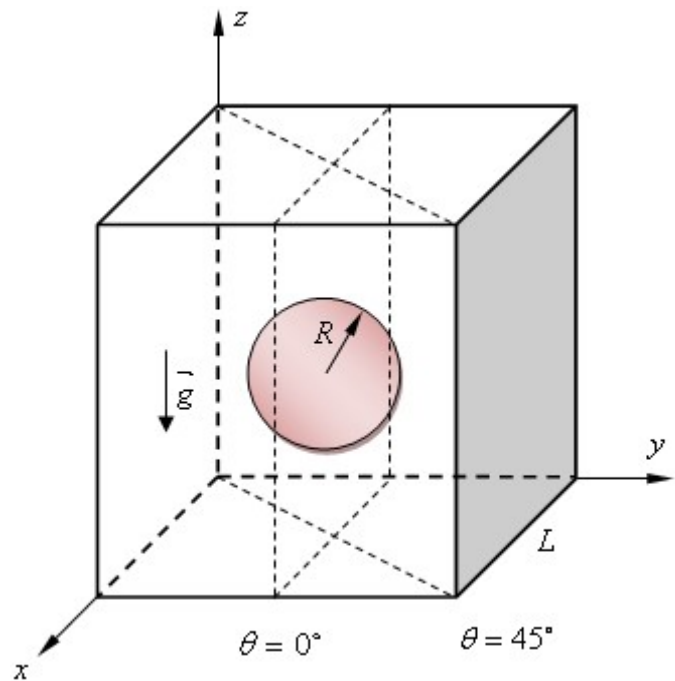

Figure 5 : Sphere in a cubic cavity. Doted lines show the two calculation planes

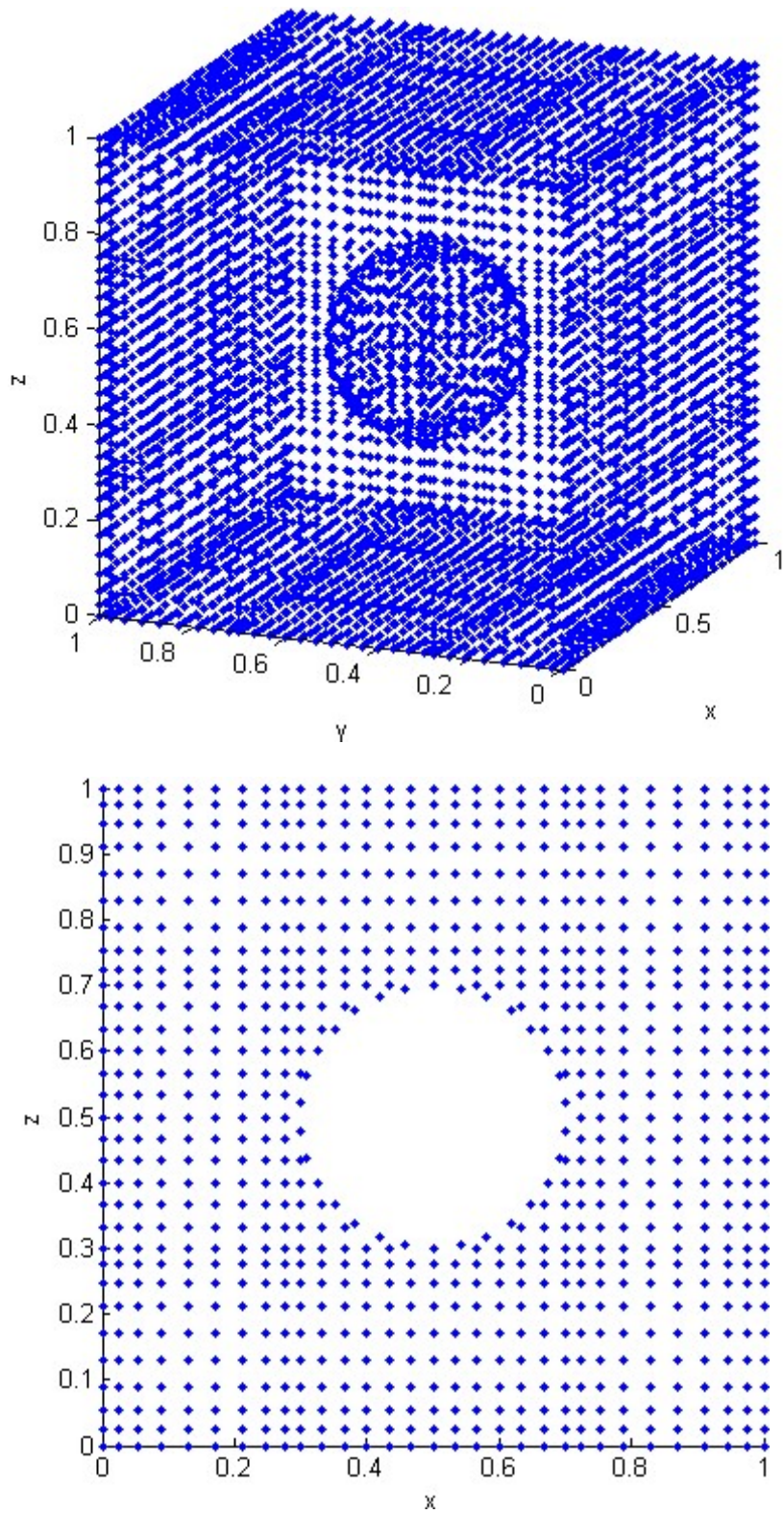

Figure 6 : Example of non uniform grid used for the simulations (31×31 on the square surface)

The general flow pattern in this configuration can be described by noting that hot fluid near the sphere will move upward until it hits the cold wall of the enclosure where the fluid direction is changed. The fluid spreads out horizontally and moves downward producing two counter-rotating rotating vortices.

The plots of the isotherms and streamlines in the planes $\theta=0$ and $\theta=45°$ are shown on Figure 7 at $Ra=10^3$. At this low Rayleigh number, the isotherms are concentric to the sphere and nearly parallel. Heat transfer occurs mainly by conduction and two low circulating vortices develop symmetrically. At $Ra=10^4$, the flow increases and the vortices centres move slightly to the upper part of the cavity as shown on Figure 8. One can also notice that the thermal gradient is higher near the bottom of the sphere.

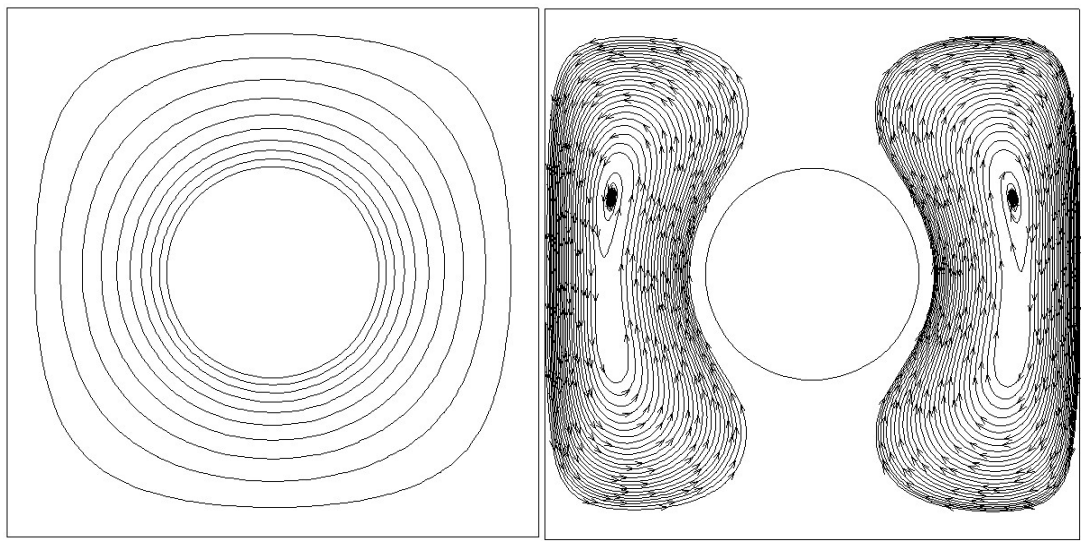

(a)

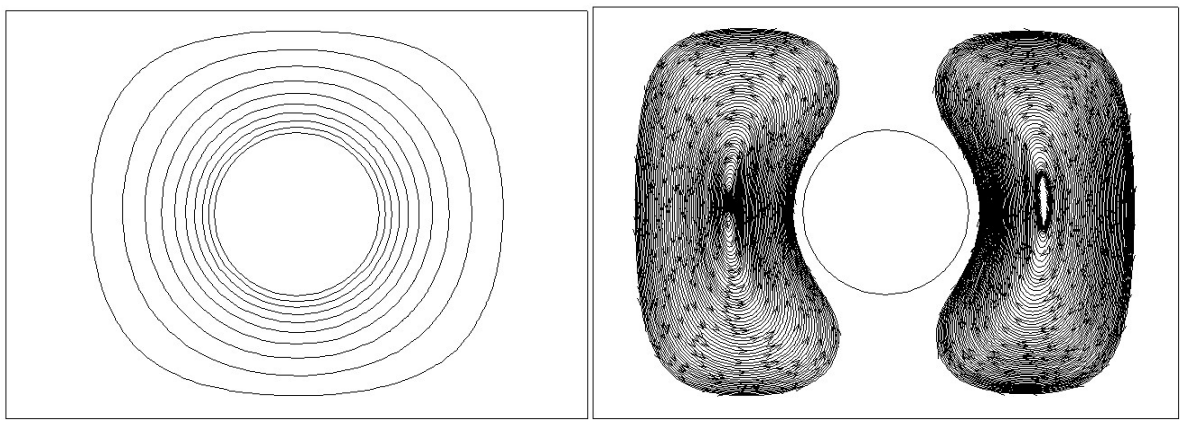

(b)

Figure 7 : Isotherms and streamlines for $\theta=0$ (a) and $\theta=45°$ (b) at $Ra=10^3$

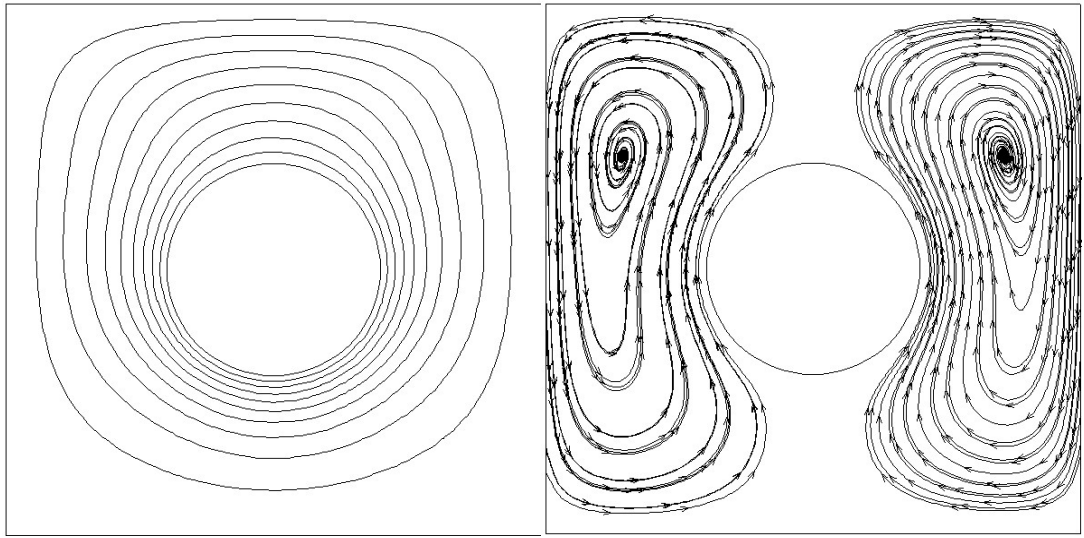

(a)

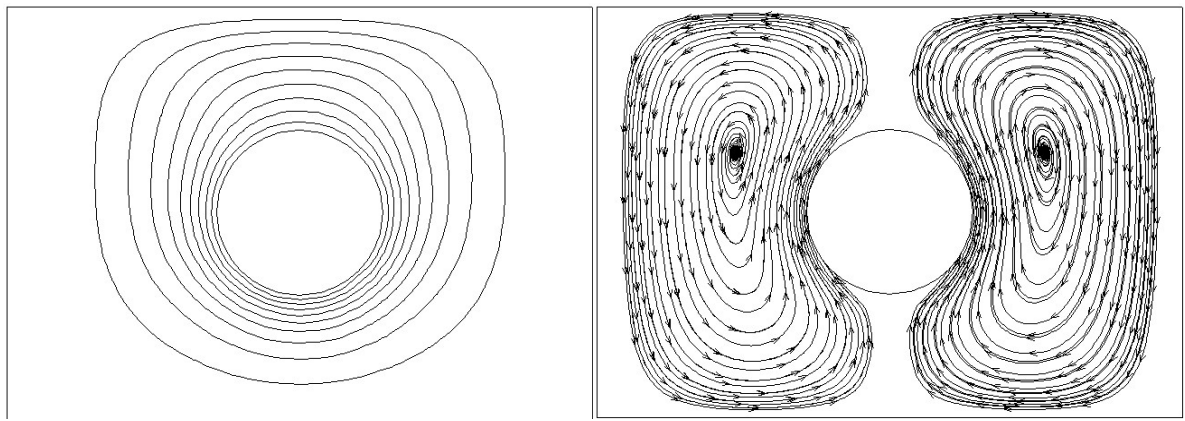

(b)

Figure 8 : Isotherms and streamlines for $\theta=0$ (a) and $\theta=45°$ (b) at $Ra=10^4$

When the Rayleigh number increases, the flow strongly hits the top of the enclosure and the isotherms move upward. In Figure 9, we display the contour plot of temperature field and the streamlines at $Ra=10^5$. It can be seen that completely different forms of the isothermal lines are obtained; they show that a thermal plume is set in the cavity. The vortices centres are now located in the upper part of the cavity where the flow is more intense than in the lower part.

A further increase of the Rayleigh number up to $10^6$ leads to a much stronger flow. Natural convection is the dominant heat transfer mechanism; the core of the two rotating vortices moves up and isotherm contours are greatly distorted.

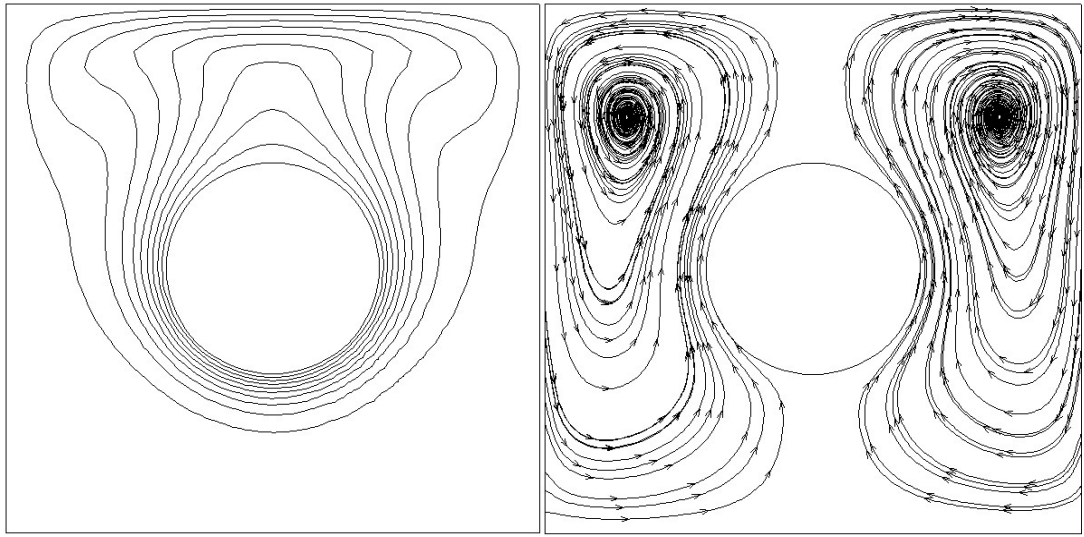

(a)

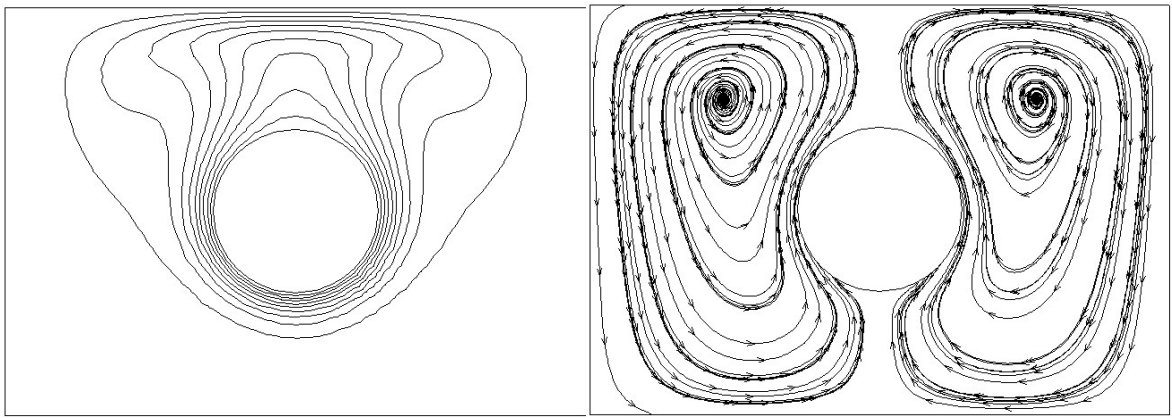

(b)

Figure 9 : Isotherms and streamlines for $\theta=0$ (a) and $\theta=45°$ (b) at $Ra=10^5$

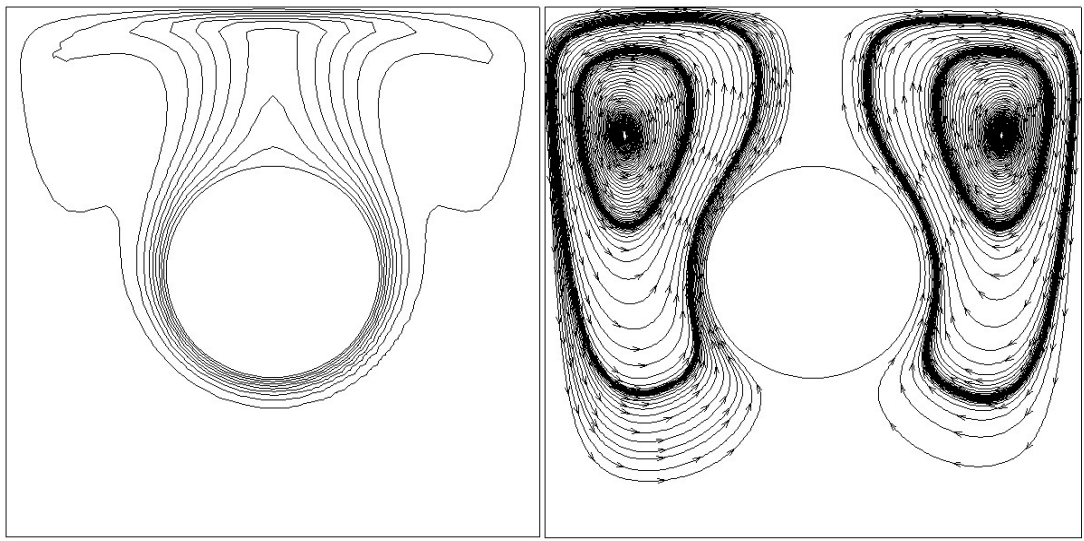

(a)

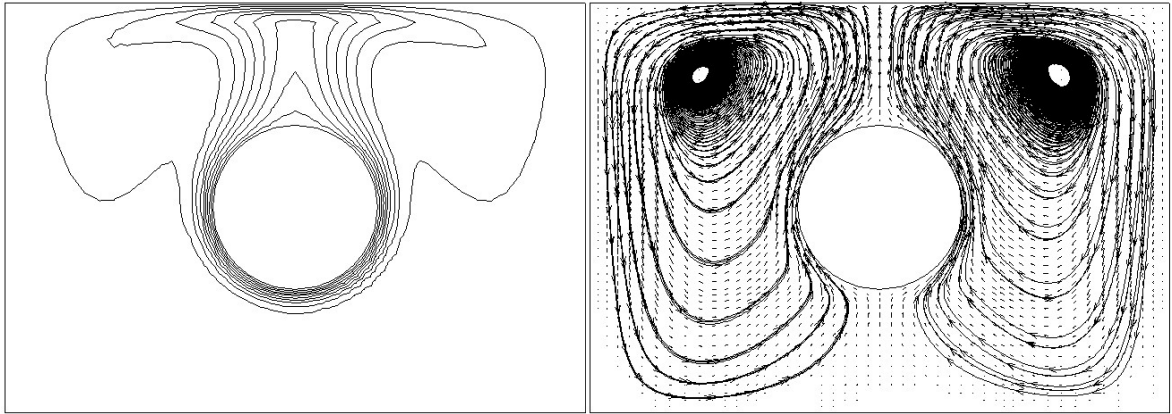

(b)

Figure 10 : Isotherms and streamlines for $\theta=0$ (a) and $\theta=45°$ at $Ra=10^6$

Finally, a more quantitative comparison is given in Figure 11 where we have plotted the local Nusselt number on the upper surface at the mid-plane. The present results are of comparable accuracy than the results of [31].

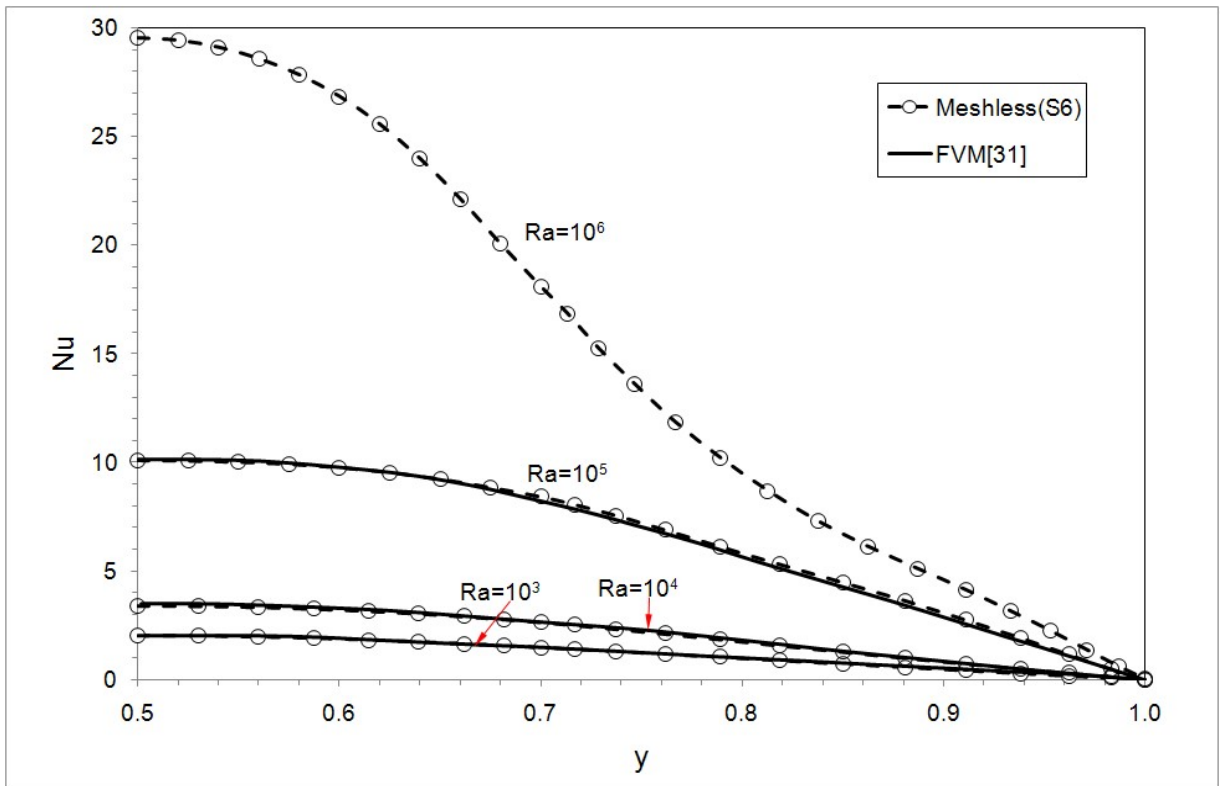

Figure 11 : Nusselt number on the top wall as a function of y at $x = 0.5$

# 6  Conclusion

In this paper we have presented a new method for solving 3-D convection in complex geometries by means of a meshless method in the vector-potential-vorticity formulation of the Navier Stokes equations. Velocity fields can be calculated from the vector potential with good accuracy. Two examples have been treated. Our calculations have shown the good accuracy of the proposed method as compared to available results from the literature. By overcoming the pressure-velocity coupling and therefore by decreasing the calculation time, this method can be used now to more evolved problems and to tackles coupled problems where other phenomena such as radiation are present.